# Should Observations be Grouped for Effective Monitoring of Multivariate Process Variability?

**Jimoh Olawale Ajadi[1] and Inez Maria Zwetsloot[2]**

Department of Systems Engineering and Engineering Management, City University of Hong Kong, Kowloon, Hong Kong

18th June 2019

**Abstract**

A multivariate dispersion control chart monitors changes in the process variability of multiple correlated quality characteristics. In this article, we investigate and compare the performance of charts designed to monitor variability based on individual and grouped multivariate observations. We compare one of the most well-known methods for monitoring individual observations- a multivariate EWMA chart proposed by Huwang et al[1] - to various charts based on grouped observations. In addition, we compare charts based on monitoring with overlapping and nonoverlapping subgroups. We recommend using charts based on overlapping subgroups when monitoring with subgroup data. The effect of subgroup size is also investigated. Steady-state average time to signal is used as performance measure. We show that monitoring methods based on individual observations are the quickest in detecting sustained shifts in the process variability. We use a simulation study to obtain our results and illustrated these with a case study.

**Keywords:** multivariate control chart, dispersion, individual observation, overlapping subgroup, non-overlapping subgroup

---

[1] Mr. Jimoh Olawale Ajadi is a PhD student in the Department of Systems Engineering and Engineering Management at City University of Hong Kong. His email address is joajadi2-c@my.cityu.edu.hk

[2] Dr. Inez Maria Zwetsloot is an assistant professor in the Department Systems Engineering and Engineering Management at City University of Hong Kong. Her email address is i.m.zwetsloot@cityu.edu.hk

## 1. Introduction

Multivariate control charts are employed to monitor several correlated quality characteristics; for example, monitoring simultaneously the strength and diameter of a tensile fiber, or monitoring the inner diameter, thickness, and length of a tube. Hotelling[2] introduced a $T^2$ control chart; this was the first multivariate control chart. Many researchers have improved on this $T^2$ control chart, to detect small and intermediate shifts in the process mean vector quicker (for example, see Crosier[3], Healy[4], Lowry et al[5]).

Similarly, some other multivariate charts have been developed to monitor multivariate process dispersion. For example see the review articles by Yeh et al[6], Bersimis et al[7] and the references therein. These charts monitor the process covariance matrix (Σ) with the objective to detect changes in Σ quickly. Many different methods exist for monitoring multivariate dispersion. Methods are based on monitoring various characterizations of the covariance matrix such as the determinant, trace, entropy or eigenvalues. We shall review the literature on methods for monitoring variability in Section 3.

Irrespective of the selected chart, one issue in setting up a multivariate chart for dispersion is how to arrange the observations. Many charts are designed for groups (samples) of observations. For example, when we observe 4 observations every hour, we can use these 4 observations as a group to monitor after the hour has passed. However, these observations may also be available at 15-minute intervals and can be used as individual samples each 15 minutes to update the control chart. Hence, we have a choice, a chart for individual observations may be applied, or one could artificially group the observations and use methods for grouped observations. The objective of this study is to provide guidelines for this choice by investigating the performance of monitoring methods based on multivariate individual and grouped observations.

The grouping of observations over time is often referred to as temporal aggregation. Recently, Zwetsloot and Woodall[8] reviewed the literature on temporal aggregation and the principle of rational subgrouping in statistical process monitoring. The principle of rational subgrouping prescribes that the observation stream is (artificially) broken up into meaningful groups based on process knowledge so as to most easily detect an assignable cause once it occurs. Zwetsloot and Woodall[8] mentioned that the selected level of temporal aggregation must not violate the principle of rational subgroup. In our paper, we assume that no rational subgrouping is needed, that is, the observations belong to a constant comparable process until an assignable cause occurs. We do form subgroups of our observations by aggregating $n$ vectors over time (temporal aggregation). These artificial groups of observations are used to monitor the process. Our reason to form groups is to possibly speed up the detection of assignable causes.

Temporal aggregation can either involve overlapping (moving window) or non-overlapping (fixed window) subgroups. For each new subgroup in an overlapping subgroup approach, the oldest observation is removed, and the newest is added to the group. For each new subgroup in a non-overlapping approach, the observations are grouped into consecutive non-overlapping subgroups. Table 1 differentiates between the individual and grouped observations by using six bivariate observations of a patient's systolic and diastolic blood pressure.

Table 1: Illustration of individual observations, overlapping subgroups and non-overlapping subgroups.

| Date | Day 1 | Day 2 | Day 3 | Day 4 | Day 5 | Day 6 |
|---|---|---|---|---|---|---|
| Individual observations | [173, 86] | [176, 87] | [163, 84] | [169, 85] | [153, 82] | [152, 83] |
| Overlapping subgroups | | [173, 86] | [176, 87] | [163, 84] | [169, 85] | [153, 82] |
| | | [176, 87] | [163, 84] | [169, 85] | [153, 82] | [152, 83] |
| Non-overlapping subgroups | | [173, 86] | | [163, 84] | | [153, 82] |
| | | [176, 87] | | [169, 85] | | [152, 83] |

Reynolds and Stoumbos[9–11] investigated whether to use individual or grouped observations for monitoring with a univariate chart. The authors considered CUSUM, EWMA and Shewhart charts. They concluded that the Shewhart chart has a better performance when $n > 1$ compared to $n = 1$ (individual observations). In addition, they showed that the best sampling choice for CUSUM and EWMA charts is to take individual observations. In contrast, Yang[12] and Wu et al[13] argued that the optimal subgroup size for the univariate CUSUM chart is 2 or 3 and for the Shewhart chart, it is 3 or 4.

All current literature on selecting an appropriate subgroup size focusses on univariate monitoring. No work exists on selecting the subgroup size for multivariate process dispersion charts. Therefore, in this article, we study whether observations should be grouped for effective monitoring of multivariate process dispersion. We will compare methods for individual monitoring to methods designed for grouped observations by applying them to the same data stream, which we artificially group for the latter charts.

We give an overview of the models and notations used in Section 2. In Section 3, we review the literature on multivariate dispersion charts and then explain the selected methods included in our comparison. We give details about the simulation procedure in Section 4 and study the effect of subgroup size for the charts based on grouped observations in Section 5. We interpret the results of our comparison study in Section 6. In Section 7, a case study is used to illustrate our findings. Conclusion and recommendations are discussed in Section 8.

## 2. Model, Assumptions and Data Aggregation

Throughout this paper, we are interested in monitoring the variability of a p-dimensional vector, $X_t$, representing p-characteristics which may be correlated. For this purpose, we assume that we observe $X_t$ at times $t = 1,2,3, ...$, each equidistant in time and that $X_t \sim N_p(\boldsymbol{\mu}, \boldsymbol{\Sigma})$, where the process vector mean and covariance matrix are denoted by $\boldsymbol{\mu}$ and $\boldsymbol{\Sigma}$, respectively. When the process is in-control, we define $\boldsymbol{\mu} = \boldsymbol{\mu_0}$ and $\boldsymbol{\Sigma} = \boldsymbol{\Sigma_0}$. In this study, we standardize $X_t$ by transforming it to $Y_t$ as

$$Y_t = \Sigma_0^{-\frac{1}{2}}(X_t - \boldsymbol{\mu}_0) \text{ for } t = 1,2,3 ...$$

Thus, $Y_t$ follows a standardized multivariate normal distribution $N(\boldsymbol{\mu}_Y, \boldsymbol{\Sigma}_Y)$, where $\boldsymbol{\mu}_Y = \Sigma_0^{-\frac{1}{2}}(\boldsymbol{\mu} - \boldsymbol{\mu}_0)$ and $\boldsymbol{\Sigma}_Y = \Sigma_0^{-\frac{1}{2}} \Sigma \Sigma_0^{-\frac{1}{2}}$. When the process is in-control, $Y_t \sim N(\boldsymbol{0}, \boldsymbol{I_p})$, where $\boldsymbol{I_p}$ is a $p \times p$ identity matrix.

The *n* period grouped observations can also be considered for process monitoring. When these subgroups are non-overlapping, they are denoted by

$$Y_T^{[n]} = [Y_{T-n+1} ... Y_{T-1} \, Y_T] \text{ , we observe } Y_T^{[n]} \text{ for } T = n, 2n, 3n, ....$$

Here $Y_T^{[n]}$ is a *p* by *n* data matrix. Note that we only observe $Y_T^{[n]}$ at time intervals which are $n$ apart; $T = n, 2n, 3n, ....$

Alternatively, we can group observations according to a moving window approach, creating overlapping subgroups. These groups are formed as

$$Y^{[n]}_{T'} = [Y_{T'-n+1} \ldots Y_{T'-1} \, Y_{T'}], \text{ we observe } Y^{[n]}_{T'} \text{ for } T' = n,\ n+1,\ n+2,\ \ldots.$$

Here $Y^{[n]}_{T'}$ is a $p$ by $n$ data matrix. Note that we observe $Y^{[n]}_{T'}$ at each time point for $T' \geq n$. After forming the first group, the oldest observation is removed and a new one is added to form the next group. This is quite different from $Y^{[n]}_{T}$, the non-overlapping groups, which are only observed at time intervals of size $n$.

Throughout our paper, we assume that the process parameters ($\mu_0$ and $\Sigma_0$) are known, for the sake of simplicity. However, in practice, the parameters need to be estimated, usually through a Phase I study. Recently, Zwetsloot and Ajadi[14] compared the performance of the most common EWMA control charts for the process dispersion. The comparison was based on effects of the Phase I estimate under normal and non-normal distributed data. For details on Phase I estimations see Jensen et al[15] and Chakraborti et al[16]. We have no reason to believe that our results will change when we also incorporate estimation.

## 3. Compared Methods for Monitoring Multivariate Dispersion

Several techniques have been developed in the literature for monitoring variability of a multivariate chart based on individual and grouped observations. We first discuss the methods based on individual observations in Section 3.1 and next, in Section 3.2, and 3.3, we discuss monitoring methods based on overlapping and non-overlapping subgroups.

### 3.1 Monitoring Dispersion Based on Individual Observations

One of the most well-known techniques for monitoring multivariate variability, based on individual observations, is by Huwang et al[1] who introduced the multivariate exponentially weighted mean squared deviation (MEWMS) and multivariate exponentially weighted moving variance (MEWMV) charts. These charts employ the trace of the estimated covariance matrix as the monitoring statistic. The authors prefer the trace to the determinant of the covariance matrix because they argued that deriving the exact distribution of the latter is more difficult.

The idea of applying trace to detect changes in the process covariance matrix is not always effective because in several out-of-control situations while some values of the diagonal elements of the estimator increase, some other elements may reduce. Therefore, Memar and Niaki[17] improved on the MEWMS and MEWMV charts by using the deviation of each diagonal element (of the covariance matrix estimator) from their expected values. Khoo and Quah[18] proposed a chart that is based on the successive difference between the pairs of multivariate observations. In addition, other methods were introduced by Yeh et al[19], Hawkins and Maboudou-tchao[20], Mason et al[21], Djauhari[22], and Fan et al[23].

The MEWMV chart is effective in detecting changes in both the process mean vector and the variability. Whereas the MEWVS chart is designed under the assumption that the process mean vector is stable when monitoring the process. For our study, we selected the MEWMS chart over MEWMV chart because we are only interested in monitoring the process variability.

To set up the MEWMS chart, we compute the multivariate exponentially weighted moving average as

$$\boldsymbol{E}_t = \omega \boldsymbol{Y}_t \boldsymbol{Y}'_t + (1-\omega)\boldsymbol{E}_{t-1}, \text{ for t=1,2,3,…}$$

where $0 < \omega < 1$ is a smoothing constant and $\boldsymbol{E_0} = \boldsymbol{Y}_1 \boldsymbol{Y}'_1$. The smoothing parameter ($\omega$) cannot be 1 so that $\boldsymbol{E}_t$ is guaranteed to be positive definite. A small value of $\omega$ is usually selected to detect small and intermediate shifts in the process. Next, we compute the trace of $\boldsymbol{E}_t$ which will be our monitoring statistic

$$tr(\boldsymbol{E}_t) = \sum_{i=1}^{p} e_{t,ii},$$

where $e_{t,ii}$ is the $i-th$ diagonal element of $\boldsymbol{E}_t$. We compare $tr(\boldsymbol{E}_t)$ to the upper and lower control limits (UCL and LCL);

$$LCL_t/UCL_t = p \pm L\sqrt{2\, p\, C_t},$$

where $C_t = \frac{\omega}{2-\omega} + \frac{2-2\omega}{2-\omega}(1-\omega)^{2(t-1)}$ and $L$ is a control chart constant. The control chart constants in Table 2 are obtained through numerical search methods using steady state performance. For details on steady state, see section 4.1. These values are slightly different from the value of $L$ reported by Huwang et al[1] because they used zero-state performance measures. We signal when $tr(\boldsymbol{E}_t) > UCL_t$ or $tr(E_t) < LCL_t$. We refer to this chart as MEWMS.

Table 2: MEWMS Control chart constants for in-control $ATS_0 = 370$

| $p$ | 2 | | 10 | |
|---|---|---|---|---|
| $\omega$ | 0.2 | 0.9 | 0.2 | 0.9 |
| $L$ | 3.4964 | 4.9 | 3.02 | 3.779 |
| $UCL^*$ | 4.3309 | 10.8644 | 14.5020 | 25.2868 |
| $LCL^*$ | -0.3309 | -6.8644 | 5.4981 | -5.2868 |

* This is a time-varying limit, here we report the value it converges to as $t \to \infty$.

### 3.2 Monitoring Variability Based on Non-overlapping Subgroups

In this subsection, we consider monitoring non-overlapping subgroup data, recall that we defined $Y_T^{[n]}$ as subgroups observed at time intervals $T = n, 2n, 3n, \dots$. Many multivariate dispersion charts have been proposed to monitor grouped observations. For example, Alt[24] proposed the generalized variance chart (GVC), the monitoring statistic for this chart is the determinant of the estimated covariance matrix. The challenge of using the determinant to monitor dispersion is that different covariance matrices can have the same determinant. Thus, Guerrero-Cusumano[25] introduced a multivariate control chart based on conditional entropy. Conditional entropy is based on the diagonal elements of the covariance matrix. Tang and Barnett[26] introduced a chart that is based on decomposing the covariance matrix into independent $\chi^2$ statistics. Yeh and Lin[27] introduced a box-chart. This chart can detect changes in the process vector mean and covariance matrices simultaneously. The box-chart uses the probability integral transformation to change statistics into the same distribution. Levinson et al[28] introduced the G-statistic for monitoring the covariance matrix (named as G chart). The G-statistic tests for equality between two covariance matrices. The authors recommended that the G chart should be used together with Hotelling's $T^2$ chart. Vargas and Lagos[29] compared the performance of different multivariate charts including a modified version of the G chart (RG) which was introduced by the authors. The authors recommended the RG chart for monitoring the process dispersion because it shows good detection probability for both increases and decreases in the covariance matrix. Djauhari et al[30] proposed a vector variance

(VV) chart. The monitoring statistic of the VV chart is the sum of the square of all elements of $\Sigma$. The chart is efficient for monitoring large datasets with high dimensions. Unlike the generalized variance chart, the VV chart can be used when the covariance matrix is singular. Abbasi et al[31] proposed a transformation based multivariate chart. The method employed a transformation technique to remove the correlation structure between the quality characteristics.

We choose the GVC proposed by Alt[24] for monitoring variability of grouped observations because of its popularity. In the GVC, the determinant of the sample covariance matrix is plotted against the control limits. Without loss of generality, we monitor the standardized data $Y_T^{[n]}$. The sample covariance matrix is computed as

$$S_T = \frac{1}{(n-1)} Y_T^{[n]} \left(Y_T^{[n]}\right)^T \text{ for } = n, 2n, 3n, \dots, \quad (1)$$

where $(\ )^T$ denotes the transpose. For the GVC, the monitoring statistic is the determinant of $S_T$:

$$\det(S_T) = |S_T|$$

We compare $\det(S_T)$ with the control limits

$$UCL_{GVC} = \left(b_1 + L_{GVC}\sqrt{b_2}\right) \quad (2)$$

and

$$LCL_{GVC} = max\{(b_1 - L_{GVC}\sqrt{b_2}), 0\}. \quad (3)$$

Here $b_1$ and $b_2$ are well-known control charting constants (for example see Montgomery[32], page 531-532). $L_{GVC}$ is the chart's constant obtained through numerical search methods using steady state performance and it is provided in Table 3. Note that because we use standardized data the factor $|\Sigma_0|$ is equal to 1 and therefor does not appear in formula's (2) and (3). A signal is obtained whenever $\det(S_T) > UCL_{GVC}$ or $\det(S_T) < LCL_{GVC}$. We refer to this chart as GVC.

*Table 3: Control Constants and LCL and UCL for the GVC defined for ATS = 370*

| $p$ | 2 | | | 10 | | |
|---|---|---|---|---|---|---|
| $n$ | 3 | 5 | 10 | 11 | 15 | 20 |
| $L_{GVC}$ | 4.778 | 3.571 | 2.550 | 0.660 | 1.375 | 1.435 |

| LCL | 0 | 0 | 0 | 0 | 0 | 0 |
|---|---|---|---|---|---|---|
| UCL | 5.8420 | 4.0302 | 2.5356 | 0.0023 | 0.0582 | 0.1864 |

Since the determinant and the trace are commonly used to summarize the overall variability of a covariance matrix, we introduce another non-overlapping chart. This chart applies the trace of the sample covariance matrix, $S_T$ as the monitoring statistic. The trace has been used in some form for monitoring the multivariate process variability, for example, see Alt and Smith[33] and it is defined as:

$$tr(\mathbf{S}_T) = \sum_{i=1}^{p} s^2_{t,ii},$$

where $s^2_{t,ii}$ is the $i-th$ diagonal element of $S_T$ as defined in equation (1). We plot $tr(\mathbf{S}_T)$ against the control limits given as:

$$LCL = \frac{\chi^2_{p(n-1),\frac{\alpha}{2}}}{n-1}, \quad (4)$$

and

$$UCL = \frac{\chi^2_{p(n-1),1-\frac{\alpha}{2}}}{n-1}. \quad (5)$$

Where $\chi^2_{v,\beta}$ represents the $\beta$ quantile of the Chi-squared distribution with $v$ degree of freedom. See the derivation of the $UCL$ and $LCL$ in the appendix; also see Section 4.1 on how to convert the average run length obtain from these limits to steady state ATS performance we used for this study. This chart is referred to as non-overlapping trace covariance chart: NTCC. We provide the appropriate type I error for the NTCC to satisfy $ATS_0 = 370$ in Table 4.

*Table 4: Type I error and LCL and UCL for the NTCC defined for ATS = 370*

| $p$ | 2 | | | 10 | | |
|---|---|---|---|---|---|---|
| $n$ | 3 | 5 | 10 | 11 | 15 | 20 |
| $\alpha = 0.0027n$ | 0.0081 | 0.0135 | 0.0270 | 0.0297 | 0.0405 | 0.0540 |
| LCL | 0.0929 | 0.3667 | 0.8203 | 7.1778 | 7.7066 | 8.1205 |
| UCL | 7.6676 | 5.2878 | 3.7502 | 13.3181 | 12.5973 | 12.0696 |

### 3.3 Monitoring Variability Based on Overlapping Subgroups

In this subsection, we consider monitoring subgroup data, where the subgroups are formed in a moving window (overlapping subgroups). Recall that we defined $Y_{T'}^{[n]}$ as subgroups observed at time intervals $T' = n, n + 1, n + 2, ....$ This approach potentially has the benefit of observing shifts quicker than non-overlapping subgroups.

Holmes and Mergen[34] introduced a mean square successive difference covariance matrix estimator. This estimator applies the overlapping subgroups technique. Sullivan and Woodall[35] compared some covariance matrices estimators based on both non-overlapping and overlapping subgroups. Their comparison shows that the estimator proposed by Holmes and Mergen[34] has the best performance. However, we have not seen any research that applies this overlapping subgroup technique in Phase II monitoring for multivariate process variability.

We include two charts in our comparison designed for overlapping subgroups. The first chart uses the sample covariance matrix ($S_{T'}$):

$$S_{T'} = \frac{1}{(n-1)} Y_{T'}^{[n]} \left(Y_{T'}^{[n]}\right)^T \text{ for } T' = n, n+1, n+2, ....$$

And uses the trace of $S_{T'}$, as monitoring statistic:

$$tr(S_{T'}) = \sum_{i=1}^{p} s_{T',ii}^2,$$

where $s_{T',ii}^2$ is the $i - th$ diagonal element of the $S_{T'}$ matrix. We signal a potential change in the process whenever $tr(S_{T'})$ exceeds either of the two control limits (UCL and LCL) in equation (4) and equation (5). Since the $tr(S_{T'})$ is affected by serial correlation due to the overlapping subgroups, the type I error ($\alpha$) cannot give the desired average time to signal, therefore, we search using numerical methods for the appropriate (symmetric) $LCL$ and $UCL$ to give steady state performance of $ATS_0 = 370$, values are displayed in Table 5. We referred to this chart as the Overlapping Trace Covariance Chart; OTCC.

*Table 5: LCL and UCL for the OTCC defined for ATS = 370*

| $p$ | 2 | | | 10 | | |
|---|---|---|---|---|---|---|
| $n$ | 3 | 5 | 10 | 11 | 15 | 20 |
| LCL | 0.056738 | 0.257927 | 0.646292 | 6.564921 | 7.120220 | 7.572200 |
| UCL | 8.746398 | 6.101049 | 4.301269 | 14.29510 | 13.45174 | 12.81552 |

The second chart we include is based on the mean squared successive difference (MSSD) covariance matrix, initially proposed by Holmes and Mergen[34], which is defined as:

$$\boldsymbol{MSSD}_{T'} = \frac{1}{2(n-1)} (\boldsymbol{Y}_{T'}^{[n]} - \boldsymbol{Y}_{T'-1}^{[n]}) (\boldsymbol{Y}_{T'}^{[n]} - \boldsymbol{Y}_{T'-1}^{[n]})^T, \text{ for } T' = n+1, n+2, n+3, \ldots$$

Sullivan and Woodall[35] mentioned that the mean successive difference covariance matrix ($\boldsymbol{MSSD}_{T'}$) is a robust estimate of the covariance matrix. We monitor based on the trace of the $\boldsymbol{MSSD}_{T'}$ estimator yielding the following monitoring statistic:

$$tr(\boldsymbol{MSSD}_{T'}) = \sum_{i=1}^{p} MSDD_{i,T'},$$

where $MSSD_{i,T'}$ is the $i-th$ diagonal element of the $\boldsymbol{MSSD}_{T'}$. We compare $tr(\boldsymbol{MSSD}_{T'})$ against the upper and lower control limits (UCL and LCL) and signal when $tr(\boldsymbol{MSSD}_{T'}) > UCL$ or $tr(\boldsymbol{MSSD}_{T'}) < LCL$. We referred to this chart as the Overlapping Trace Mean successive differences Chart; OTMC. The $UCL$ and $LCL$ are derived in the appendix. We also search for $LCL$ and $UCL$ to satisfy steady state performance of $ATS_0 = 370$ for the OTMC and provide the values in Table 6.

*Table 6: LCL and UCL for the OTMC defined for ATS = 370*

| $p$ | 2 | | | 10 | | |
|---|---|---|---|---|---|---|
| $n$ | 3 | 5 | 10 | 11 | 15 | 20 |
| LCL | 0.036211 | 0.185008 | 0.517543 | 6.047177 | 6.650895 | 7.156457 |
| UCL | 9.726786 | 6.872069 | 4.823367 | 15.20931 | 14.19955 | 13.42693 |

Table 7 provides an overview of the selected charts for our comparison of monitoring methods based on individual observations or grouped observations (either overlapping or non-overlapping).

Table 7: Overview of the compared charts for monitoring multivariate dispersion

| Aggregation Level | Charts | Descriptions | Variants considered |
|---|---|---|---|
| Individual observations | MEWMS | Multivariate exponentially weighted mean square proposed by Huwang et al[1] | $\omega = 0.2$ and $\omega = 0.9$ |
| Non-overlapping subgroups of size $n$ | GVC | Generalized variance chart proposed by Alt[24] | $n = 3,5,10$ when $p = 2$ and $n = 11,15,20$ when $p = 10$ |
| | NTCC | Non-overlapping chart based on the trace of sample covariance matrix | $n = 3,5,10$ when $p = 2$ and $n = 11,15,20$ when $p = 10$ |
| Overlapping subgroups of size $n$ | OTCC | Overlapping chart based on the trace of sample covariance matrix | $n = 3,5,10$ when $p = 2$ and $n = 11,15,20$ when $p = 10$ |
| | OTMC | Overlapping chart based on the trace of mean successive differences estimator. | $n = 3,5,10$ when $p = 2$ and $n = 11,15,20$ when $p = 10$ |

## 4. Simulation Study and Performance Criterium

In this section, we discuss the simulation procedure and the performance measure we employ for the comparison of the selected methods (Table 7).

### 4.1 Performance Metrics

We applied Average Time to Signal (ATS) as the performance metrics. ATS is the average time between the start of a shift and the first signal. We denote $ATS_0$ and $ATS_1$ as the in-control and out-of-control $ATS$ values, respectively. Using ATS as a performance measure is preferred to average run length (expected number of plotted statistics before the signal is detected) because the aggregation rate for each of the compared charts is different. Average run length (ARL) is related to ATS when $n$ has regularly spaced time interval according to

$$\text{Individual monitoring: } ATS = ARL$$

$$\text{Non-overlapping subgroup: } ATS = n * ARL$$

$$\text{Overlapping subgroup: } ATS = ARL + n - 1$$

We assume a sustained shift in the process can start at a random moment in time, $ie.$ also within a subgroup. This is referred to as steady-state performance. Steady-state ATS is employed as the performance measure for the compared methods. The mathematical relationship between the zero state ($ATS_{zerostate}$) and steady state ($ATS_{steadystate}$) ATS is given as: $ATS_{steadystate} = ATS_{zerostate} + \frac{n}{2}$, where $n$ is the subgroup size.

### 4.2 Simulation Study

In Section 6, we will compare the performance of the selected charts. We consider $p = 2$ (bivariate observations) and $p = 10$, and use steady-state ATS as the performance measure. 50000 Monte Carlo simulation runs are employed to evaluate the ATS for each method. Our in-control model is $\mu_0 = (0,0,\ldots,0)_{1 \times p}$ and

$$\Sigma_0 = \begin{pmatrix} \sigma_1^2 & 0 & \cdots & 0 \\ 0 & \sigma_2^2 & \cdots & 0 \\ \vdots & & \ddots & \vdots \\ 0 & 0 & \cdots & \sigma_p^2 \end{pmatrix}_{p \times p}, \qquad (6)$$

and we represent the shifts in the process variance as

$$\Sigma_1 = \delta \Sigma_0, \qquad (7)$$

where $\delta$ is the shift in the process. Another out-of-control scenario is when both $\delta$ and $\rho$ increase where we consider overall shifts in the process as:

$$\Sigma_2 = \delta \begin{pmatrix} \sigma_1^2 & \rho\sigma_1\sigma_2 & \cdots & \rho\sigma_1\sigma_{p-1} & \rho\sigma_1\sigma_p \\ \rho\sigma_1\sigma_2 & \sigma_2^2 & \cdots & \rho\sigma_2\sigma_{p-1} & \rho\sigma_2\sigma_p \\ \vdots & & \ddots & & \vdots \\ \rho\sigma_1\sigma_p & \rho\sigma_2\sigma_p & \cdots & & \rho\sigma_{p-1}\sigma_{p-2}\sigma_p^2 \end{pmatrix}_{p \times p}, \qquad (8)$$

for $\rho = 0.2, 0.6,$ and $0.8$. In addition, we consider the partial shifts in the process where the first $q$ variables are shifted and the rest are unchanged.

$$\Sigma_3 = \begin{pmatrix} \sigma_1^2 & \rho\sigma_1\sigma_2 & \cdots & \rho\sigma_1\sigma_q & 0 & 0 & \cdots & 0 \\ \rho\sigma_1\sigma_2 & \sigma_2^2 & \cdots & \rho\sigma_2\sigma_q & 0 & 0 & \cdots & 0 \\ \vdots & \vdots & \ddots & \vdots & \vdots & \vdots & \ddots & \vdots \\ \rho\sigma_1\sigma_q & \rho\sigma_2\sigma_q & \cdots & \sigma_q^2 & 0 & 0 & \cdots & 0 \\ 0 & 0 & \cdots & 0 & \sigma_{q+1}^2 & 0 & 0 & 0 \\ 0 & 0 & \cdots & \vdots & 0 & \sigma_{q+2}^2 & 0 & 0 \\ \vdots & \vdots & \ddots & \vdots & \vdots & \vdots & \ddots & \vdots \\ 0 & 0 & \cdots & 0 & 0 & 0 & \cdots & \sigma_p^2 \end{pmatrix}.$$

(9)

## 5. Effects of Subgroup Size

First, we study the charts based on subgroups in order to understand the effect of the subgroups size $n$. We run 50000 Monte Carlo simulations for each chart. For the $p = 2$ scenario, we consider subgroups of size $n = 3,\ 5$ and 10, and for $p = 10$, we consider $n = 11, 15$ and 20.

Figure 1 displays the ATS curves of each chart for various subgroup sizes. Charts are designed with $ATS_0 = 370$. For the out-of-control case in Figure 1, we assume an overall shift in the process variances as in equation (7). Note that ATS values in Figure 1 and 2 are on a log-scale.

First consider $p = 2$ in the subgraphs on the left, we observed from Figure 1a and e that for small and moderate shifts ($\delta < 2.5$) in the process, the GVC and OTCC perform best when they are based on large subgroups size. However, these charts are more effective in detecting large shifts in the process when they are based on intermediate subgroup sizes of $n = 5$. As shown in Figure 1c, the NTCC performs similar in terms of $ATS$ at small and moderate shifts in the process for both subgroup size $n = 5$ and $n = 10$. For large shifts ($\delta > 2.5$), subgroup size of $n = 3$ and $n = 5$ yield the quicker detection. In Figure 1g we see that the OTMC perform similar irrespective of the subgroup size for all shifts in the process.

Next consider $p = 10$, in Figure 1b, d, f, h, we see that the $ATS_1$ values, for each of the charts and for $p = 10$ is similar when the shifts in the process variability is small. However, when there are intermediate and large shifts in the process, small subgroup sizes yield lower $ATS$ values.

Next consider NTCC (Figure 1c and d) and OTCC (Figure 1e and f), the only difference between these charts is the formation of subgroups through overlapping or non-overlapping subgroups. It is evident that both NTCC and OTCC have similar performance when the shifts in the process is very small. However, for moderate and large shifts in the process, we observe that the OTCC shows a better performance than the non-overlapping chart (NTCC) irrespective of the subgroup size and for each of the $p$ levels. It is also evident that OTCC is the quickest in detecting shifts among the charts based on monitoring with grouped observations when $p = 2$ or $p = 10$. Consequently, we recommend to use charts based on overlapping subgroups if we decide to monitor with subgroup data.

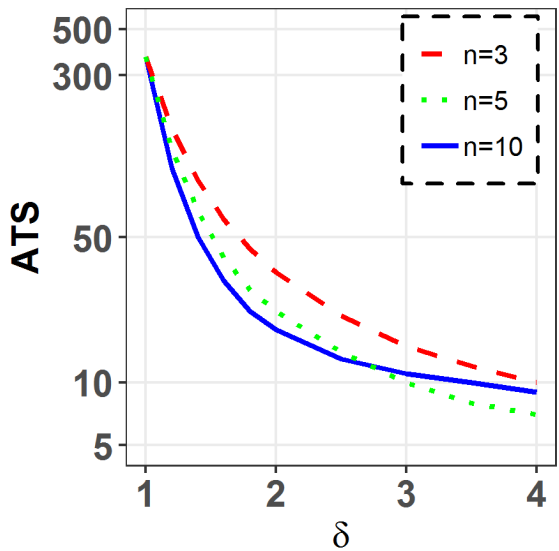

(a): GVC for $p = 2$

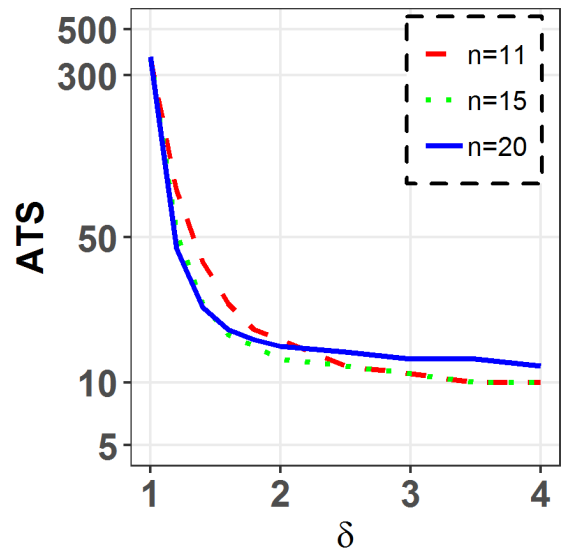

(b): GVC for $p = 10$

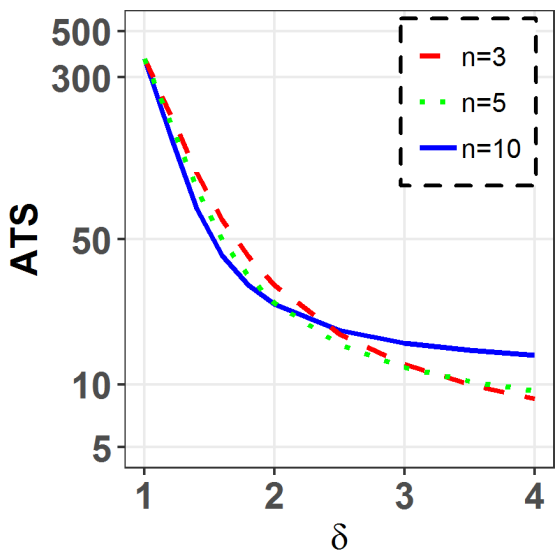

(c): NTCC when $p = 2$

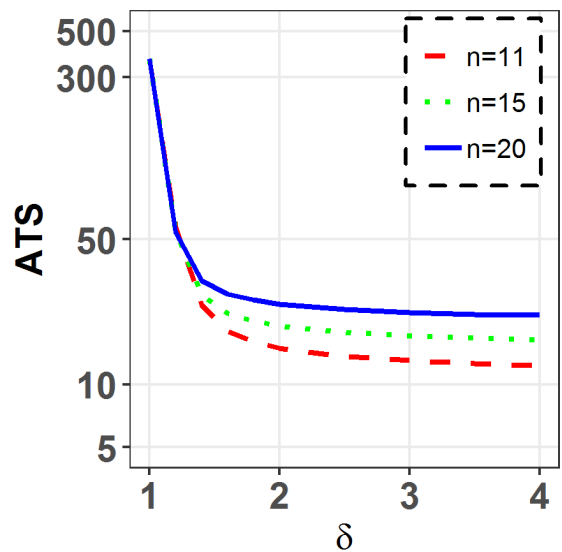

(d): NTCC when $p = 10$

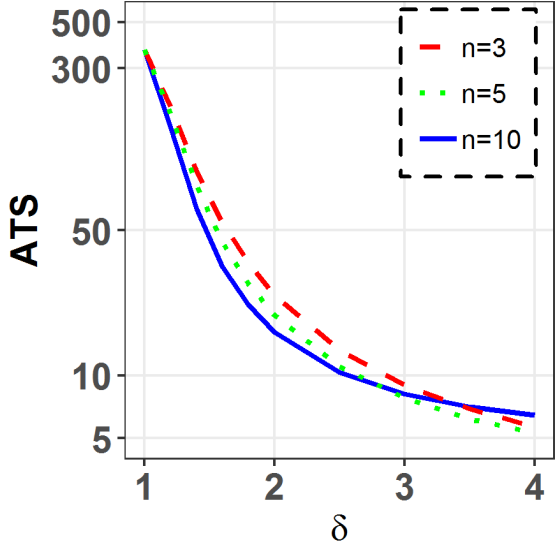
(e): OTCC when $p = 2$

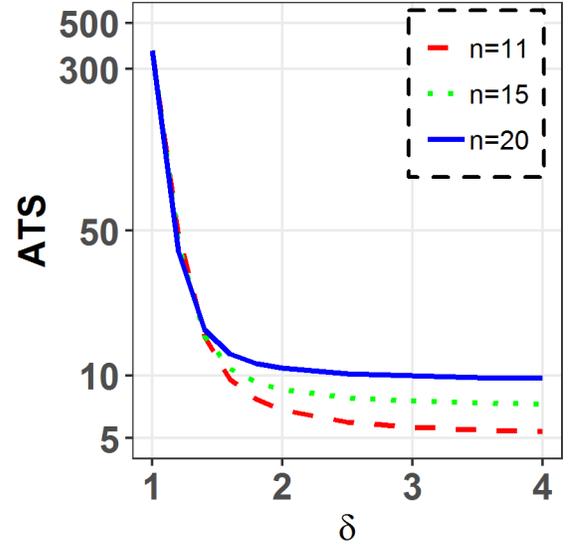
(f): OTCC when $p = 10$

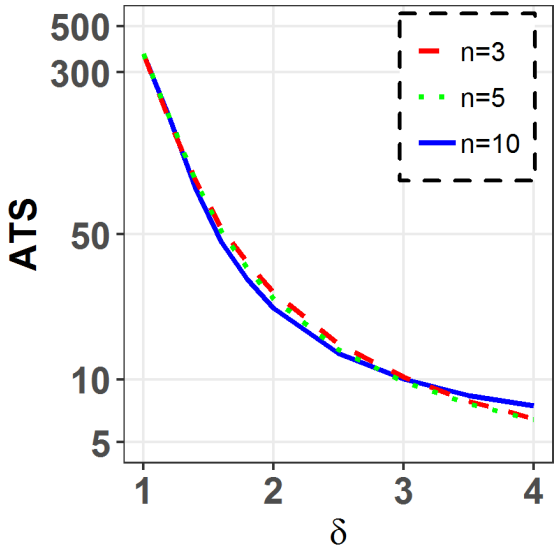
(g): OTMC when $p = 2$

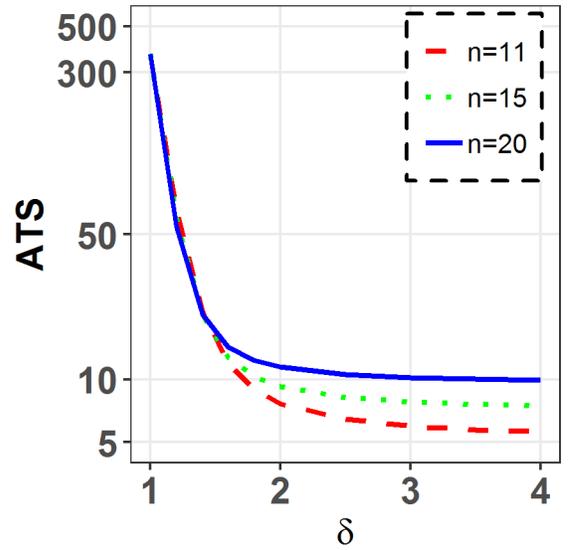
(h): OTMC when $p = 10$

**Figure 1:** ATS curves for various charts under out-of-control data where $\Sigma_1 = \delta\Sigma_0$ and various samples sizes $n$. Note for $\delta = 1$ the data are in-control and the charts are tuned for an $ATS_0 = 370$.

Next, we considered the scenario where there is a shift in the correlation coefficient as defined in Eq(8), we set $\rho = 0.6$; results are displayed in Figure 2. Note that this implies that the data are out-of-control even if $\delta = 1$. We see that GVC does not perform very well (Figure 2a & b). Its ATS value even increase for small values of $\delta$. Also, the NTCC perform similarly at small and moderate shifts irrespective of the subgroup size but it has its best performance at large shift for small and intermediate subgroup size. We observe from Figure 2 (e-h) that the OTCC and OTMC show similar $ATS_1$ values at different subgroup sizes ($n$) and $p$'s.

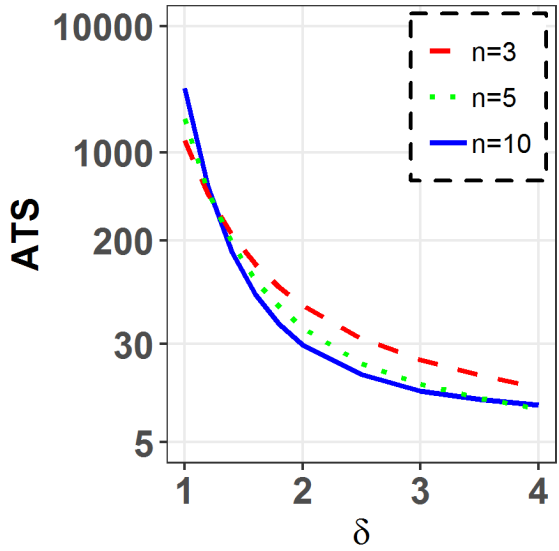

(a): GVC for $p = 2$

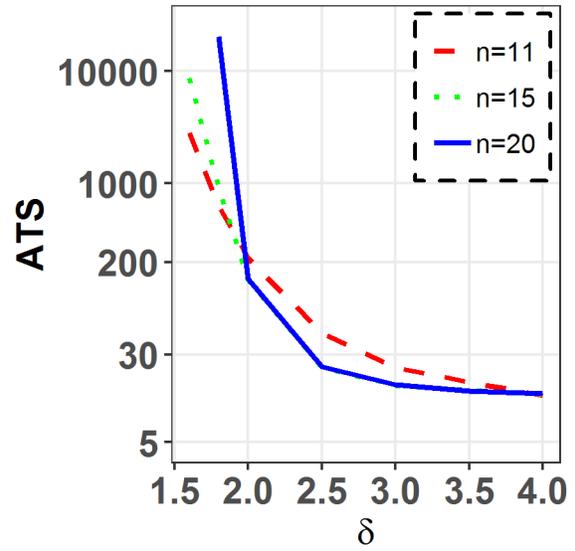

(b): GVC for $p = 10$

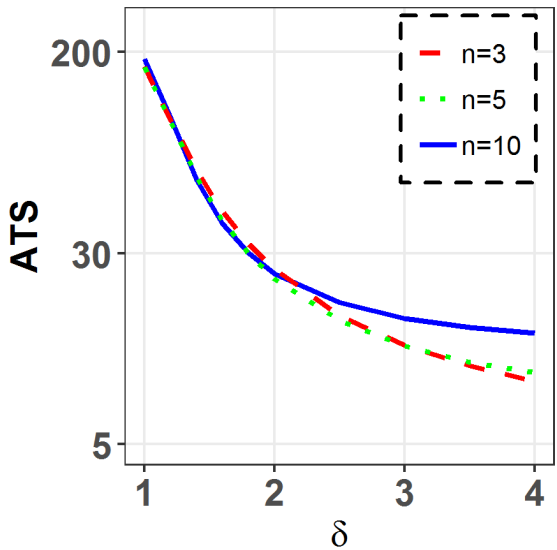

(c): NTCC when $p = 2$

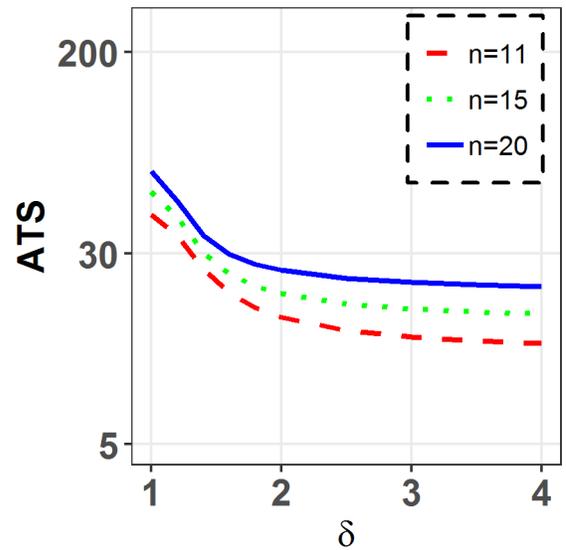

(d): NTCC when $p = 10$

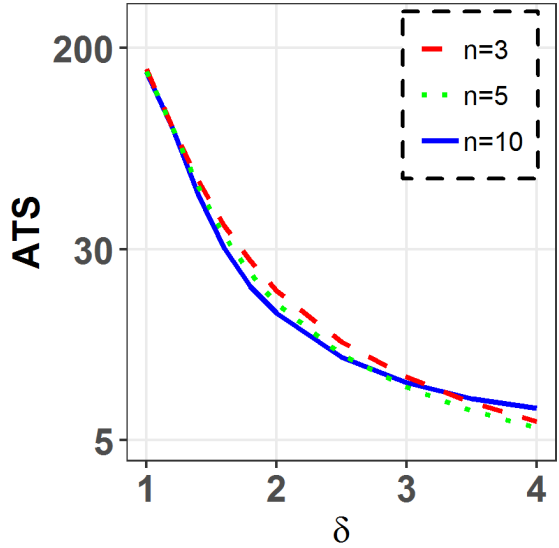
(e): OTCC when $p = 2$

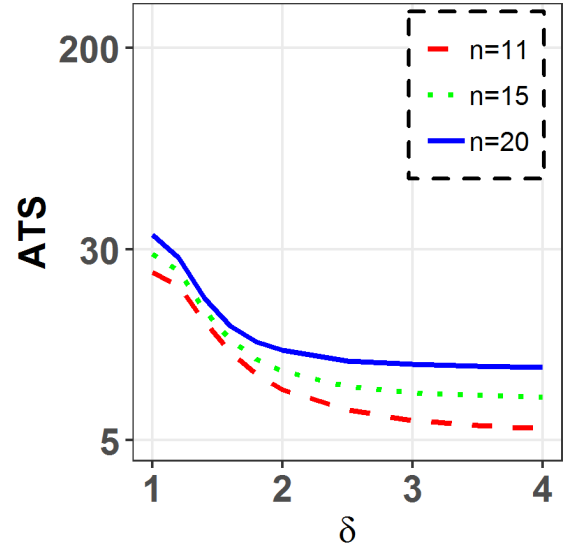
(f): OTCC when $p = 10$

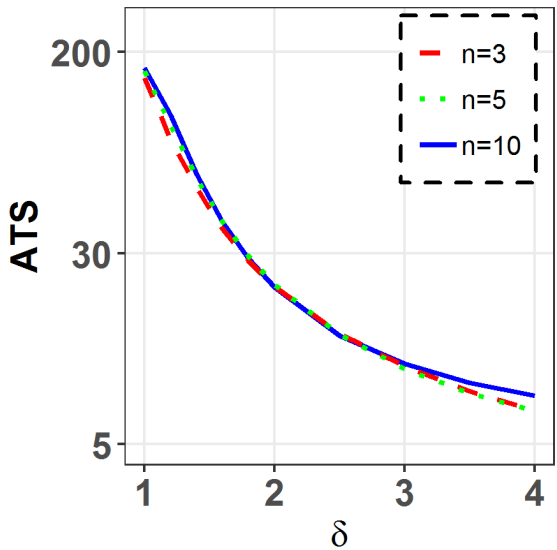
(g): OTMC when $p = 2$

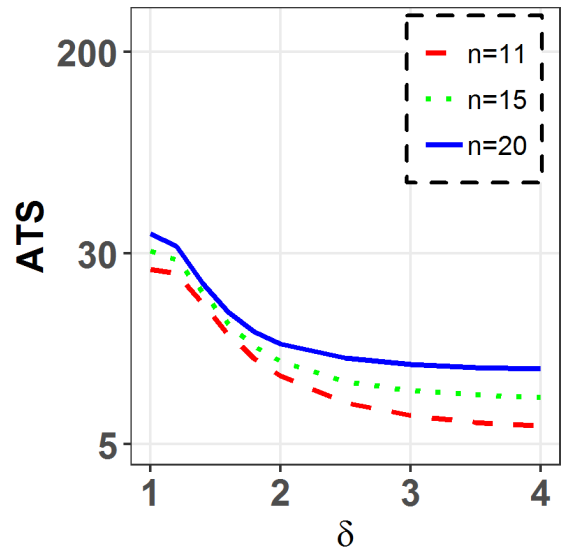
(h): OTMC when $p = 10$

**Figure 2:** ATS values for various charts under out of control data form for $\Sigma_2$ and $\rho = 0.6$ for p=2 and p=10 at various samples sizes $n$.

When the value of $p$ is high, we recommend monitoring with a small subgroup size. If the subgroup size is large, there may be a delay in receiving an alarm because the signal is detected after the group is formed. In general, we recommend a small subgroup size irrespective of the value of $p$ except if we are interested in detecting small and moderate shifts in the process variances and the $\rho = 0$ for GVC, NTCC and OTCC's when $p = 2$.

## 6. Performance Comparison

In this section, we compare the performance of the charts based on monitoring the process variability of the individual observations, nonoverlapping and overlapping subgroups. This comparison is based on out-of-control ATS values when $ATS_0 = 370$. In the previous section, we saw that the charts based on grouped observations perform best for small and moderate shifts in the process when $n = 10$ for $p = 2$ and for $n = 11$ for $p = 10$. Consequently, we used a subgroup size of 10 and 11 in the following comparison. Results are displayed in Tables 8-11 where grey highlighted values correspond to in-control performance and bold face indicates the best performing chart, *ie*. lowest ATS value.

For the out-of-control scenario in Tables 8 and 10, we simulated an overall increase and decrease in the process variances as well as a shift in the correlation coefficients ($\rho$=0.2, 0.6, and 0.8) as defined in equation (8). In Table 9 and 11, we considered partial shifts in the process variance and/or $\rho$ as in equation (9) with $q = 1$ for $p = 2$ and $q = 3$ for $p = 10$. We denote the values of $ATS > 10000$ with $*$ in Tables 8 through 11.

In Tables 8 and 9, we considered $p = 2$, and noticed that the OTCC performs best among the compared charts when the process variance(s) decreases. We also observe that the OTCC has better performance than both OTMC and NTCC at all shifts in the process variability (with the exception in Table 9 where NTCC is better than both OTCC and OTMC at $\delta = 1.2$ and $\delta = 1.4$ for $\rho = 0$). GVC has a better performance than the NTCC, OTCC and OTMC when $\rho = 0$ and $\delta > 1$. The MEWMS chart has the best performance among the compared charts when the $\delta > 1$. GVC performs poorly when $\rho$ increases. Likewise, the ATS values of the MEWMS chart are also high for decreases in the process variances because the control limits of the chart are assumed to be symmetric; and thus, it fails to detect downward changes in process.

**Table 8:** ATS values for various charts with an overall process shifts for $p = 2$ where the out of control data are from equation (8). Note for $\delta = 1$ and $\rho = 0$, the data are in-control, and indicated in grey tone and the charts are tuned for an $ATS_0 = 370$.

| $\rho$ | $\delta$ | OTCC $n=10$ | OTMC $n=10$ | NTCC $n=10$ | GVC $n=10$ | MEWMS $\omega = 0.2$ | MEWMS $\omega = 0.9$ | $\rho$ | $\delta$ | OTCC $n=10$ | OTMC $n=10$ | NTCC $n=10$ | GVC $n=10$ | MEWMS $\omega = 0.2$ | MEWMS $\omega = 0.9$ |
|---|---|---|---|---|---|---|---|---|---|---|---|---|---|---|---|
| | 0.6 | **56** | 78 | 65 | * | * | * | | 0.6 | **40** | 54 | 53 | * | * | 1982 |
| | 0.7 | **103** | 138 | 119 | * | * | 4829 | | 0.7 | **66** | 86 | 84 | * | 2508 | 758 |

| ρ | δ | OTCC n=10 | OTMC n=10 | NTCC n=10 | GVC n=10 | MEWMS ω=0.2 | MEWMS ω=0.9 | ρ | δ | OTCC n=10 | OTMC n=10 | NTCC n=10 | GVC n=10 | MEWMS ω=0.2 | MEWMS ω=0.9 |
|---|---|---|---|---|---|---|---|---|---|---|---|---|---|---|---|
| 0 | 0.8 | **189** | 236 | 215 | 3228 | 3778 | 1662 | 0.6 | 0.8 | **105** | 128 | 133 | * | 767 | 360 |
|  | 0.9 | **314** | 351 | 343 | 942 | 1004 | 720 |  | 0.9 | **147** | 167 | 180 | * | 315 | 202 |
|  | 1 | 367 | 371 | 371 | 372 | 371 | 370 |  | 1 | 159 | 172 | 187 | 3208 | 160 | **130** |
|  | 1.2 | 156 | 181 | 157 | 106 | **95** | 134 |  | 1.2 | 94 | 111 | 108 | 522 | **61** | 64 |
|  | 1.4 | 63 | 82 | 70 | 49 | **41** | 66 |  | 1.4 | 50 | 63 | 60 | 161 | **33** | 38 |
|  | 1.6 | 33 | 46 | 42 | 30 | **24** | 38 |  | 1.6 | 30 | 41 | 40 | 74 | **22** | 26 |
|  | 2 | 16 | 22 | 24 | 17 | **12** | 18 |  | 2 | 17 | 22 | 25 | 30 | **12** | 15 |
|  | 2.5 | 10 | 13 | 18 | 12 | **8** | 10 |  | 2.5 | 11 | 14 | 19 | 17 | **8** | 9 |
|  | 3.5 | 7 | 8 | 15 | 10 | **5** | **5** |  | 3.5 | 7 | 9 | 15 | 11 | **5** | **5** |
|  | 4 | 6 | 7 | 14 | 10 | **4** | **4** |  | 4 | 7 | 8 | 14 | 11 | **4** | **4** |
| 0.2 | 0.6 | **53** | 75 | 64 | * | * | * | 0.8 | 0.6 | **33** | 41 | 47 | * | 5242 | 1052 |
|  | 0.7 | **98** | 133 | 114 | * | * | 3063 |  | 0.7 | **48** | 60 | 69 | * | 1254 | 435 |
|  | 0.8 | **177** | 221 | 205 | 4119 | 2693 | 1183 |  | 0.8 | **69** | 83 | 98 | * | 452 | 229 |
|  | 0.9 | **288** | 318 | 315 | 1157 | 811 | 557 |  | 0.9 | **90** | 101 | 122 | * | 206 | 135 |
|  | 1 | 328 | 335 | 335 | 438 | 316 | **299** |  | 1 | 97 | 106 | 127 | * | 116 | **90** |
|  | 1.2 | 146 | 170 | 150 | 119 | **89** | 118 |  | 1.2 | 69 | 81 | 86 | * | 50 | **48** |
|  | 1.4 | 61 | 79 | 69 | 54 | **40** | 61 |  | 1.4 | 42 | 53 | 54 | 1880 | **29** | 30 |
|  | 1.6 | 33 | 46 | 42 | 32 | **24** | 36 |  | 1.6 | 28 | 37 | 38 | 523 | **20** | 21 |
|  | 2 | 16 | 22 | 24 | 18 | **13** | 18 |  | 2 | 16 | 22 | 25 | 107 | **12** | 13 |
|  | 2.5 | 10 | 13 | 18 | 13 | **8** | 10 |  | 2.5 | 11 | 14 | 19 | 37 | **8** | 9 |
|  | 3.5 | 7 | 8 | 15 | 10 | **5** | **5** |  | 3.5 | 8 | 9 | 15 | 16 | **5** | **5** |
|  | 4 | 7 | 8 | 14 | 10 | **4** | **4** |  | 4 | 7 | 8 | 14 | 13 | 5 | **4** |

**Table 9:** ATS values for various charts with a single process shift for $p = 2$ where the out of control data are from equation (9). Note for $\delta = 1$ and $\rho = 0$, the data are in-control, and indicated in grey tone and the charts are tuned for an $ATS_0 = 370$.

| ρ | δ | OTCC n=10 | OTMC n=10 | NTCC n=10 | GVC n=10 | MEWMS ω=0.2 | MEWMS ω=0.9 | ρ | δ | OTCC n=10 | OTMC n=10 | NTCC n=10 | GVC n=10 | MEWMS ω=0.2 | MEWMS ω=0.9 |
|---|---|---|---|---|---|---|---|---|---|---|---|---|---|---|---|
| 0 | 0.6 | **171** | 213 | 199 | 4702 | 2321 | 1027 | 0.6 | 0.6 | **99** | 122 | 125 | * | 694 | 336 |
|  | 0.7 | **236** | 278 | 266 | 1965 | 1490 | 836 |  | 0.7 | **124** | 146 | 155 | * | 455 | 258 |
|  | 0.8 | **303** | 341 | 336 | 996 | 927 | 662 |  | 0.8 | **146** | 165 | 178 | * | 310 | 202 |
|  | 0.9 | **361** | 373 | 377 | 582 | 579 | 494 |  | 0.9 | **159** | 174 | 191 | 5967 | 217 | 162 |
|  | 1 | 369 | 369 | 372 | 373 | 368 | 368 |  | 1 | 160 | 171 | 187 | 3210 | 160 | **129** |
|  | 1.2 | 260 | 273 | 254 | 191 | **167** | 205 |  | 1.2 | 129 | 144 | 147 | 1211 | 93 | **87** |
|  | 1.4 | 149 | 174 | 153 | 117 | **91** | 123 |  | 1.4 | 93 | 108 | 107 | 581 | **60** | 62 |
|  | 1.6 | 89 | 113 | 97 | 80 | **56** | 80 |  | 1.6 | 66 | 81 | 77 | 335 | **43** | 47 |
|  | 2 | 43 | 57 | 51 | 48 | **29** | 41 |  | 2 | 37 | 49 | 47 | 150 | **26** | 29 |

|   |     | OTCC | OTMC | NTCC | GVC | MEWMS | MEWMS |     |     | OTCC | OTMC | NTCC | GVC | MEWMS | MEWMS |
|---|-----|------|------|------|-----|-------|-------|-----|-----|------|------|------|-----|-------|-------|
|   | 2.5 | 24 | 32 | 32 | 24 | **12** | 16 |     | 3   | 23 | 30 | 32 | 50 | **12** | 14 |
|   | 3.5 | 13 | 22 | 21 | 20 | **10** | 12 |     | 3.5 | 13 | 17 | 21 | 36 | **10** | **11** |
|   | 4   | 11 | 17 | 19 | 17 | **8**  | 9  |     | 4   | 11 | 14 | 19 | 29 | **8**  | 9  |
|   | 0.6 | **160** | 200 | 187 | 6071 | 1850 | 834 |     | 0.6 | **68** | 79 | 96 | * | 154 | 221 |
|   | 0.7 | **219** | 258 | 252 | 2449 | 1181 | 663 |     | 0.7 | **80** | 91 | 111 | * | 292 | 172 |
|   | 0.8 | **281** | 311 | 309 | 1220 | 758 | 521 |     | 0.8 | **89** | 100 | 122 | * | 212 | 135 |
|   | 0.9 | **323** | 340 | 342 | 702 | 483 | 394 |     | 0.9 | **95** | 106 | 127 | * | 152 | 109 |
|   | 1   | 328 | 336 | 338 | 446 | 318 | 297 |     | 1   | 97 | 106 | 126 | * | 117 | **90** |
|   | 1.2 | 234 | 251 | 235 | 220 | **151** | 176 |     | 1.2 | 86 | 96 | 109 | * | 73 | **63** |
|   | 1.4 | 139 | 163 | 145 | 132 | **85** | 110 |     | 1.4 | 69 | 80 | 85 | * | 50 | **47** |
|   | 1.6 | 86 | 107 | 94 | 90 | **54** | 74 |     | 1.6 | 53 | 64 | 67 | 5833 | **37** | **37** |
| 0.2 | 2 | 42 | 55 | 50 | 52 | **29** | 40 | 0.8 | 2 | 34 | 43 | 44 | 1669 | **24** | 25 |
|   | 2.5 | 24 | 32 | 32 | 34 | **17** | 23 |     | 2.5 | 22 | 29 | 31 | 550 | **16** | 17 |
|   | 3.5 | 13 | 17 | 21 | 21 | **10** | **11** |     | 3.5 | 13 | 17 | 22 | 153 | **10** | **10** |
|   | 4   | 11 | 14 | 19 | 18 | **8** | 9 |     | 4   | 11 | 14 | 19 | 97 | **8** | **8** |

Next consider $p = 10$ in Tables 10 and 11, we notice that the GVC performs worst among the competing charts for all shifts in the process variability. The MEWMS chart ($\omega = 0.2$) performs best among the compared charts for an increase in the variances of the process variability. For a decrease in the process variances, the OTCC and OTMC's have the best performance among compared charts. In addition, the OTCC has the best performance among the charts based on monitoring with grouped observations.

Table 8 through 11 showed that the MEWMS chart ($\omega = 0.9$) has best performance when the variance is in-control ($\delta = 1$) and $\rho$ is increased. In addition, since the variability of a trace increases as $\rho$ increases, the time to signal of the control charts reduces. Thus, as expected the $ATS$ values of the charts based on using trace decreases as $\rho$ increases and when $\delta = 1$.

**Table 10:** ATS values for various charts with an overall process shifts for $p = 10$ where the out of control data are from equation (8). Note for $\delta = 1$ and $\rho = 0$, the data are in-control and the charts are tuned for an $ATS_0 = 370$.

|   |   | OTCC | OTMC | NTCC | GVC | MEWMS | MEWMS |   |   | OTCC | OTMC | NTCC | GVC | MEWMS | MEWMS |
|---|---|------|------|------|-----|-------|-------|---|---|------|------|------|-----|-------|-------|
| $\rho$ | $\delta$ | $n=11$ | $n=11$ | $n=11$ | $n=11$ | $\omega=0.2$ | $\omega=0.9$ | $\rho$ | $\delta$ | $n=11$ | $n=11$ | $n=11$ | $n=11$ | $\omega=0.2$ | $\omega=0.9$ |
|   | 0.6 | **10** | 13 | 19 | * | 18 | * |   | 0.6 | **10** | 12 | 20 | * | 12 | 95 |
|   | 0.7 | **17** | 24 | 26 | * | 50 | * |   | 0.7 | **13** | 15 | 24 | * | 16 | 52 |
|   | 0.8 | **43** | 61 | 53 | 6437 | 245 | 5267 |   | 0.8 | **17** | 19 | 31 | * | 22 | 34 |
|   | 0.9 | **147** | 184 | 164 | 1209 | 1245 | 1181 |   | 0.9 | **21** | 23 | 39 | * | 26 | 23 |
|   | 1.0 | 371 | 369 | 371 | 370 | 372 | 371 |   | 1.0 | 24  | 26  | 43  | *   | 24  | **18** |
|   | 1.2 | 48  | 67  | 58  | 85  | **34** | 72  |   | 1.2 | 21  | 25  | 36  | *   | 16  | **11** |
|   | 1.4 | 15  | 21  | 24  | 38  | **12** | 25  |   | 1.4 | 15  | 19  | 26  | *   | 10  | **8** |
|   | 1.6 | 10  | 12  | 18  | 24  | **7** | 11  |   | 1.6 | 11  | 14  | 21  | 2825 | 8   | **6** |
| 0 | 2.0 | 7   | 8   | 15  | 15  | **4** | 5   | 0.6 | 2.0 | 8   | 10  | 17  | 217 | **5** | **4** |
|   | 2.5 | 6   | 6   | 14  | 12  | **3** | 2   |   | 2.5 | 7   | 7   | 15  | 47  | **4** | 3   |
|   | 3.5 | 5   | 6   | 13  | 10  | **2** | 1   |   | 3.5 | 6   | 6   | 13  | 17  | **2** | 2   |
|   | 4.0 | 5   | 6   | 12  | 10  | **2** | 1   |   | 4.0 | 6   | 6   | 13  | 13  | **2** | 2   |

**Table 11:** ATS values for various charts with a partial process shifts ($\delta\sigma_i^2, i = 1,2,3$) for $p = 10$ where the out of control data are from equation (9). Note for $\delta = 1$ and $\rho = 0$, the data are in-control and the charts are tuned for an $ATS_0 = 370$.

| | | OTCC | OTMC | NTCC | GVC | MEWMS | MEWMS | | | OTCC | OTMC | NTCC | GVC | MEWMS | MEWMS |
|---|---|------|------|------|-----|-------|-------|---|---|------|------|------|-----|-------|-------|
| $\rho$ | $\delta$ | $n=11$ | $n=11$ | $n=11$ | $n=11$ | $\omega=0.2$ | $\omega=0.9$ | $\rho$ | $\delta$ | $n=11$ | $n=11$ | $n=11$ | $n=11$ | $\omega=0.2$ | $\omega=0.9$ |
|   | 0.6 | **105** | 135 | 121 | 2268 | 860 | 1132 |   | 0.6 | **89** | 114 | 109 | 1041 | 706 | 1323 |
|   | 0.7 | **159** | 196 | 180 | 1236 | 1129 | 887 |   | 0.7 | **124** | 149 | 150 | 603 | 1028 | 1036 |
|   | 0.8 | **243** | 274 | 261 | 762 | 965 | 673 |   | 0.8 | **167** | 188 | 199 | 393 | 1103 | 838 |
|   | 0.9 | **331** | 347 | 344 | 517 | 619 | 501 |   | 0.9 | **200** | 210 | 232 | 278 | 837 | 595 |
|   | 1.0 | 371 | 368 | 368 | 368 | 370 | 370 |   | 1.0 | 201 | 204 | 229 | 1185 | 187 | **144** |
|   | 1.2 | 225 | 250 | 228 | 221 | **141** | 201 |   | 1.2 | 128 | 143 | 150 | 624 | 83 | **79** |
|   | 1.4 | 104 | 132 | 114 | 150 | **67** | 114 |   | 1.4 | 71 | 87 | 88 | 382 | **45** | 49 |
|   | 1.6 | 55 | 75 | 66 | 111 | **37** | 69 |   | 1.6 | 44 | 56 | 57 | 259 | **29** | 33 |
| 0 | 2.0 | 24 | 32 | 33 | 72 | **17** | 31 | 0.6 | 2.0 | 23 | 30 | 33 | 146 | **16** | 19 |
|   | 2.5 | 13 | 18 | 22 | 49 | **10** | 15 |   | 2.5 | 14 | 18 | 23 | 90 | **10** | 11 |
|   | 3.5 | 8 | 10 | 17 | 38 | **7** | 9 |   | 3.5 | 9 | 11 | 17 | 64 | **7** | 8 |
|   | 4.0 | 7 | 9 | 16 | 27 | **5** | 5 |   | 4.0 | 8 | 9 | 16 | 41 | **5** | 5 |

In conclusion, MEWMS chart shows the best performance irrespective of the number of correlated quality variables for an increase in the process variances. Therefore, we recommend the practitioner to employ the multivariate chart based on monitoring with the individual observations for detecting increases in the process variability. In contrary, OTCC and OTMC's perform best among the compared charts for the decrease in the process variances. Also, we observed that the overlapping charts outperform the nonoverlapping charts. Thus, when using subgroups, one should use overlapping groups. We expect that a different design of the $LCL$ of the MEWMS chart will improve its performance for decreases in $\delta$.

# 7. Real Life Example

In this section, we apply each of the compared charts to monitor the variability of a bivariate datasets from an industrial process. The dataset is available from Santos-Fernandez[36] as well as in the MSQC package in R, see Santos-Fernandez[37]. Though the author does not provide the background information of the datasets, it was mentioned that the indust1 and indust2 were collected at different time. We combine both indust1 and indust2 into one subgroup and compute the process vector mean ($\mu$) and sample covariance matrix ($\mathbf{S}_t$) as Phase I estimates with $\mathbf{S}_t$ as in equation (1). We obtain $\mu = [4.04954 \quad 7.08866]$ and $\mathbf{S}_t = \begin{pmatrix} 0.0819 & 0.0668 \\ 0.0668 & 0.1809 \end{pmatrix}$. Based on these Phase I estimates, we simulated 100 observations where the first 50 observations are in-control, the next 30 observations have an increase in the process variability of $\delta = 2.5$ and the last 20 observations have a decrease in the process of $\delta = 0.5$. The subgroup size we use for the monitoring is $n = 5$. Huwang et al.[1] also used this approach of simulating Phase II datasets using the estimated parameters from a case study datasets.

Figure 3a shows the data, where the first 50 observations are statistically in control and Figure 3c-h show the charts. When there is an increase in the process variability, Figure 3c-h reveal that the OTCC detects 10 signals at observations 59-61,64-68, 73 and 77, the OTMC detects 12 signals at observations 59-61,64-68 and 74-77. The NTCC receives 5 signals at observations 61-65 and GVC detects 20 signals at observations 56-65 and 71-80. The MEWMS chart with $\omega = 0.9$ triggers at observations 64, 65, 69 and 74. In addition, the MEWMS chart with $\omega = 0.2$ detects 25 signals at observations 57-81.

However, when there is a decrease in the process variability, only the overlapping charts detects out-of-control signals where OTCC triggers at observations 95-98 and OTMC signals at observations 97-99.

Overall, the MEWMS chart (when $\omega = 0.2$) and the overlapping charts detect the shifts quickest after the 7 observations of increasing the variability in the process by $\delta = 2.5$ and 15 observations of decreasing the process variability by $\delta = 0.5$ respectively. This also supports the simulation results in Section 6 where the MEWMS and the overlapping charts are more effective for the increase and the decrease in the process variability respectively than the other compared charts.

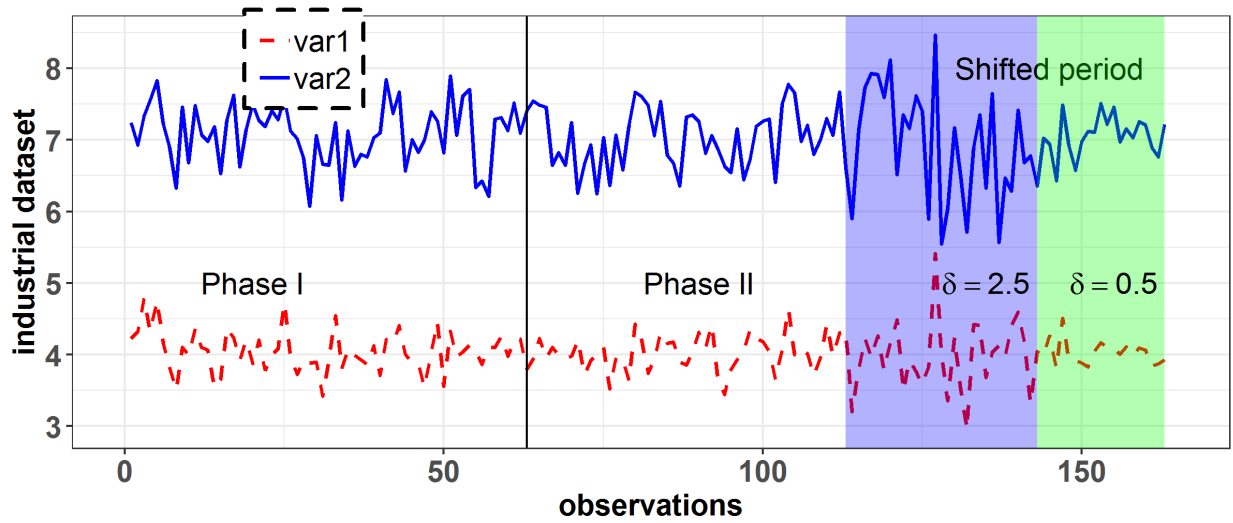
(a) Phase I and II Datasets of an Industrial Process

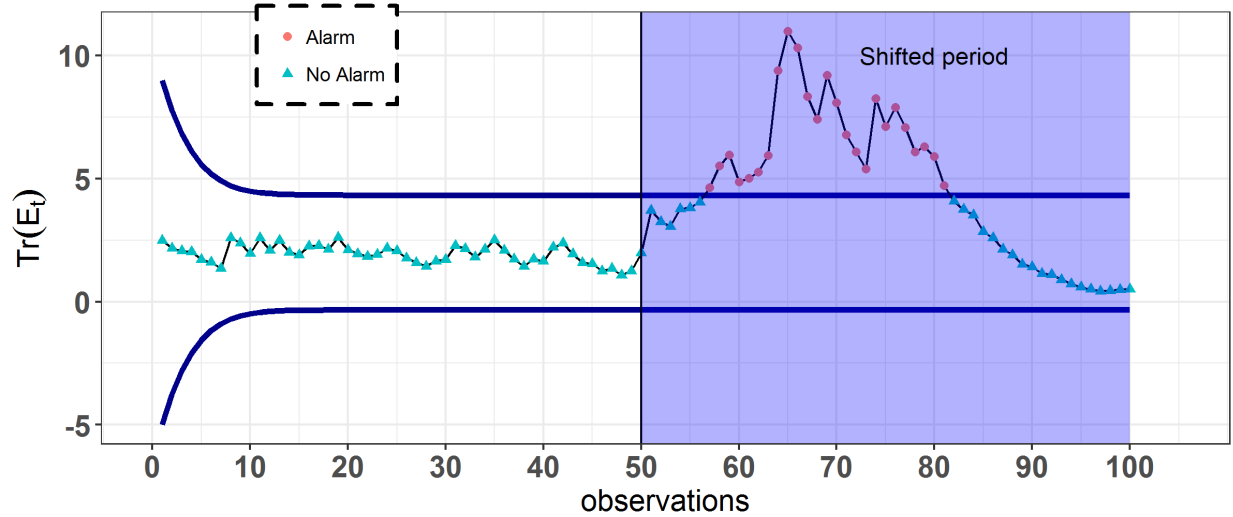
(c): MEWMS control chart when $\omega = 0.2$

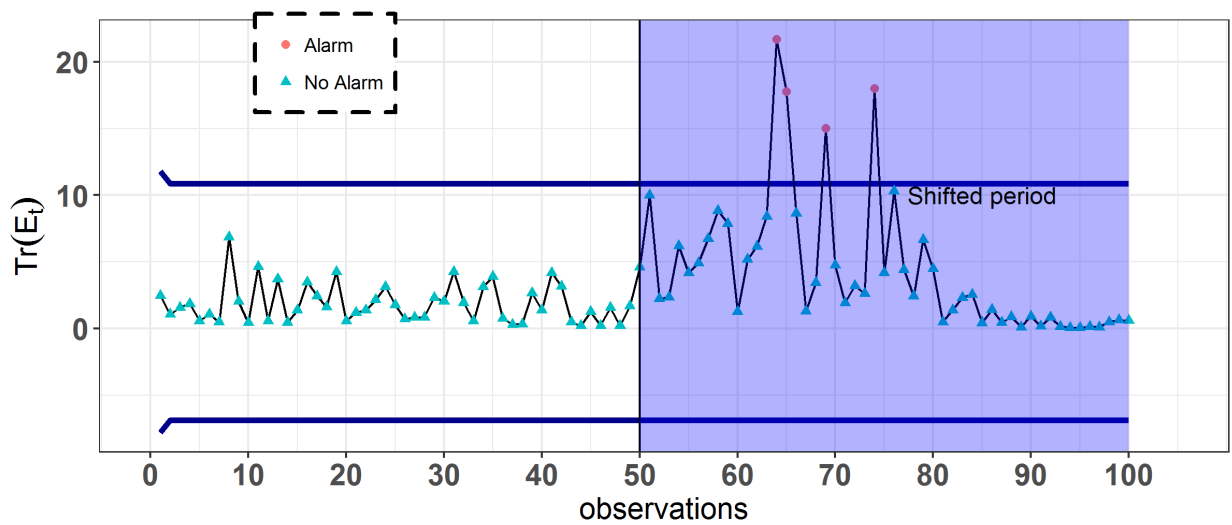
(d): MEWMS control chart when $\omega = 0.9$

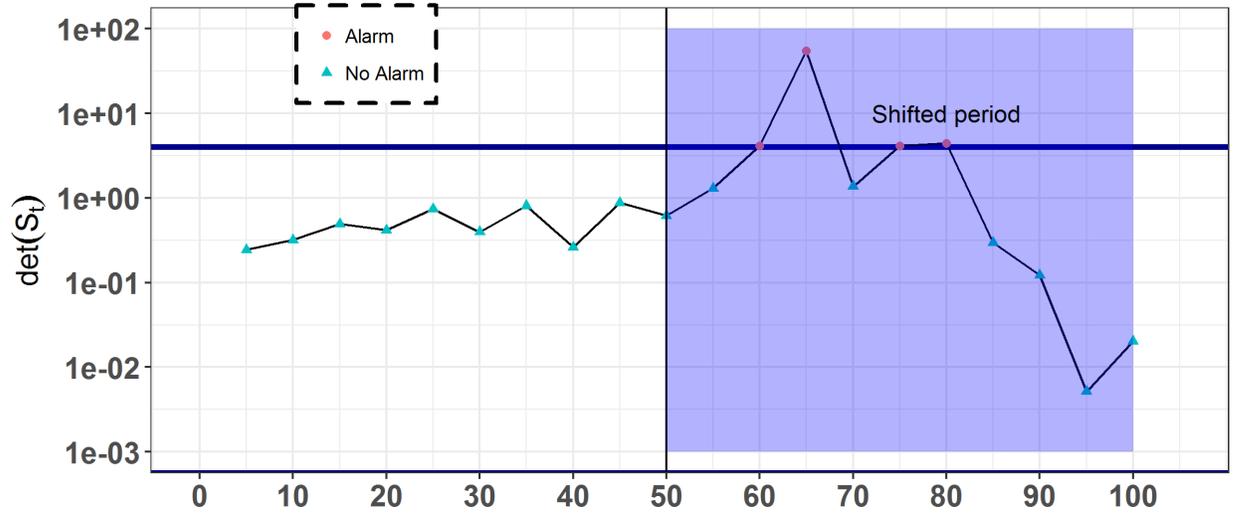

(e): Generalized variance chart

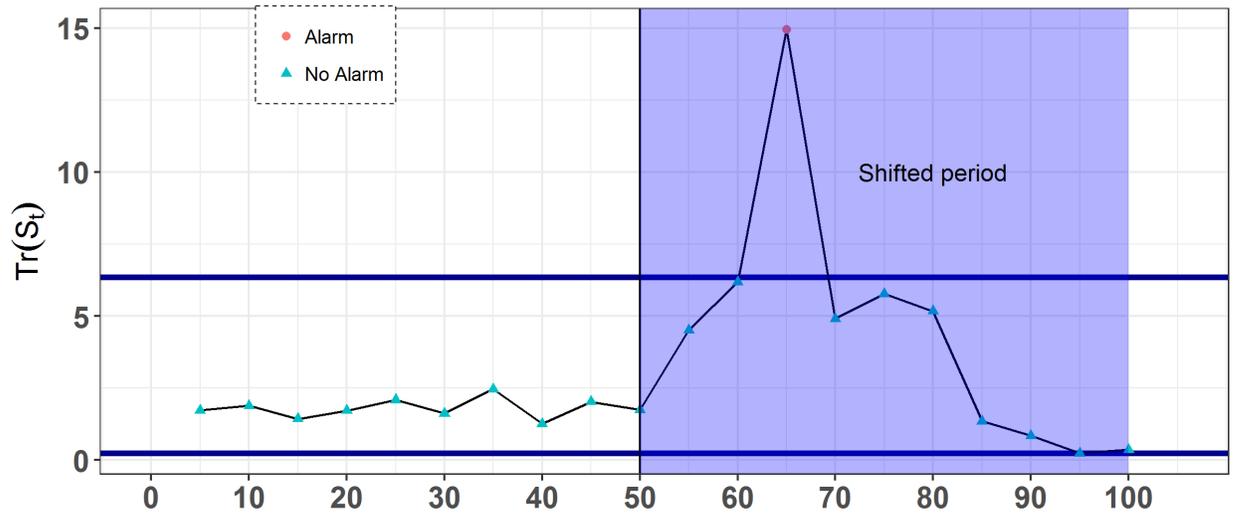

(f): Non-overlapping trace covariance chart

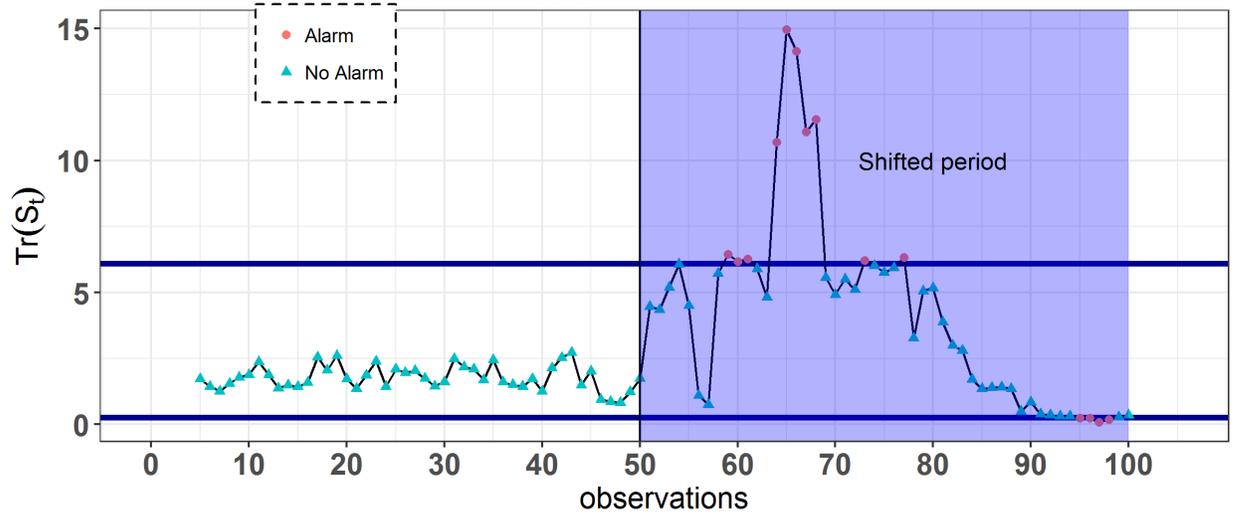

(g): Overlapping trace covariance chart

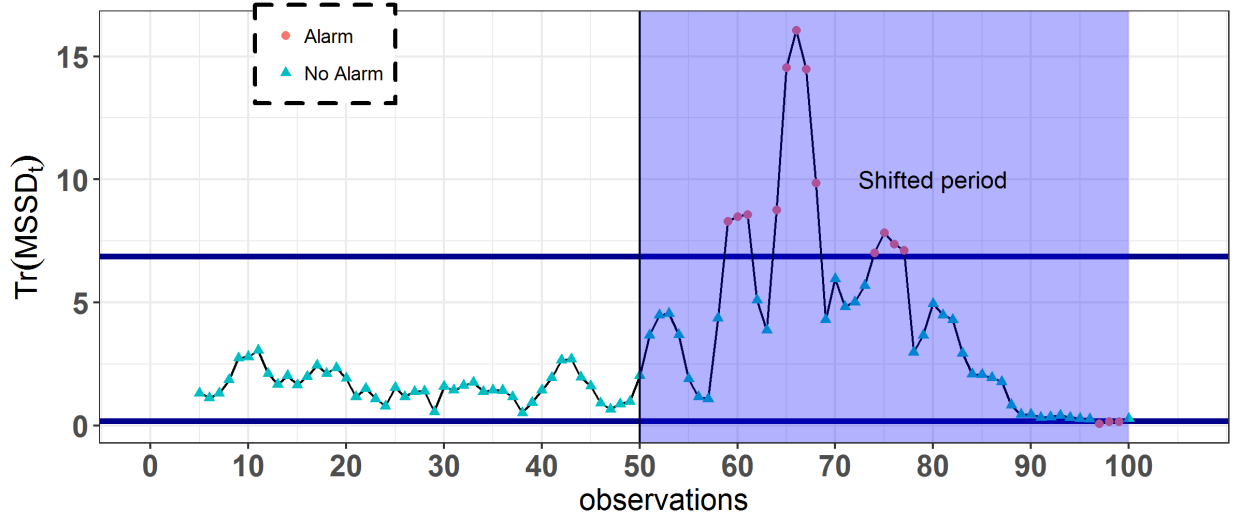

(h): Overlapping trace mean successive difference chart

**Figure 3**: Monitoring the industrial process dataset with the charts based on individual and grouped observations designed for $ATS_0 = 370$.

## 8. Conclusion, Discussion and Recommendation

In this paper, we compared the performance of multivariate charts for monitoring variability and considered individual and grouped observations. For the charts based on monitoring with individual observations, we used the MEWMS control chart proposed by Huwang et al[1]. In addition, we considered methods for monitoring with nonoverlapping and overlapping subgroups. We employed the GVC as the non-overlapping chart and we introduced the NTCC. We also developed two charts, OTCC and OTMC, for monitoring overlapping observations.

It was illustrated that both the OTCC and OTMC have the best performance when the process variances decreases. However, MEWMS chart performs best for an increase in the process variances.

In Section 5, we studied the effect of subgroup size on the charts based on monitoring with grouped observations. We found that when the subgroup size increases, the chart's performance goes down, especially for detecting large shifts. The only exception is when we monitor with OTCC, GVC and NTCC at a small and intermediate shift in the process variability for $p = 2$ and $\rho = 0$ where using large subgroup size is more effective.

We recommend that the practitioner should consider monitoring with a chart based on individual multivariate observations because consistently, it shows the best performance irrespective of the number correlated quality characteristics. This is also in agreement with the conclusion from the Reynolds and Stoumbos[9–11] for the univariate charts. In addition, we also recommend using overlapping groups when monitoring with subgroups.

In this article, we considered only $p = 2$ and $p = 10$, we expect that our conclusion on these choices would also hold for other values of $p$. We applied only three subgroups to check the effect of sample sizes on the charts based on monitoring grouped multivariate observations. The actual sample size to yield the optimal performance was not studied in our article but we recommend that this can be studied for future research.

**Acknowledgement**

The work of Inez M. Zwetsloot was partially supported by a grant from City University of Hong Kong (project 7005090).

**Appendix**

In this paper, we consider a chart based on the trace of the variance-covariance matrix. In this appendix, we derive the control limits for these charts; NTCC, OTCC and OTMC.

First consider the NTCC and recall that its monitoring statistic is equal to $Tr(S_T)$ where $S_T$ is the sample covariance matrix;

$$S_T = \begin{pmatrix} s_1^2 & s_{12} & \cdots & s_{1p} \\ s_{21} & s_2^2 & & s_{2p} \\ \vdots & & \ddots & \vdots \\ s_{p1} & s_{p2} & \cdots & s_p^2 \end{pmatrix}.$$

In order to derive the distribution of $Tr(S_T)$, we are only interested in the diagonal element $s_i^2$. Since $(n-1)s_i^2$ follows a chi-squared distribution with $n-1$ degrees of freedom it follows that $(n-1)\sum_{t=1}^{p} s_i^2$ follows a chi-squared distribution with $(n-1)p$ degrees of freedom. And as $Tr(S_T) = \sum_{t=1}^{p} s_i^2$, it follows that:

$$(n-1)Tr(S_T) \sim \chi_{p(n-1)}^2$$

hence

$$Tr(S_T) \sim \chi_{p(n-1)}^2/(n-1).$$

Therefore, the control limits of the NTCC are equal to $UCL = \dfrac{\chi_{p(n-1),1-\frac{\alpha}{2}}^2}{n-1}$ and $LCL = \dfrac{\chi_{p(n-1),\frac{\alpha}{2}}^2}{n-1}$.

Next consider the OTCC, the control limits are similar to the limits of the NTCC. However, because the $tr(\boldsymbol{S_T})$ is affected by serial correlation due to the overlapping subgroups we cannot set the type I error ($\alpha$) equal to the reciprocal of the ARL. Therefore, we employ a numerical search method to obtain desired in-control performance.

Last, we derive the control limits for the OTMC, this chart is based on the trace of the MSSD covariance matrix:

$$tr(MSSD_T) = Trace \begin{pmatrix} s_1^2 & s_{12} & \cdots & s_{1p} \\ s_{21} & s_2^2 & & s_{2p} \\ \vdots & & \ddots & \vdots \\ s_{p1} & s_{p2} & \cdots & s_p^2 \end{pmatrix} = \sum_{i=1}^{p} s_i^2,$$

where $s_i^2$ are the diagonal elements of the $MSSD_T$ matrix and they are defined as $s_i^2 = \sum_{i=2}^{n} \dfrac{(y_i-y_{i-1})^2}{2(n-1)}$. We know that $y_i - y_{i-1} \sim N(0,2)$ because $y_i \sim N(0,1)$ under in-control data and therefore $\left(\dfrac{y_i-y_{i-1}}{\sqrt{2}}\right)^2 \sim \chi_1^2$.

Hence $(n-1)s_i^2 = \sum_{i=2}^{n}\left(\dfrac{y_i-y_{i-1}}{\sqrt{2}}\right)^2 \sim \chi_{(n-1)}^2$, since $(n-1)s_i^2$ follows a chi-squared distribution with $n-1$ degrees of freedom it follows that $(n-1)\sum_{t=1}^{p} s_i^2$ follows a chi-squared distribution with $(n-1)p$ degrees of freedom.

$$(n-1)Tr(MSSD_t) \sim \chi_{p(n-1)}^2$$

The control limits of the OTMC are $UCL = \frac{\chi^2_{p(n-1), 1-\frac{\alpha}{2}}}{n-1}$ and $LCL = \frac{\chi^2_{p(n-1), \frac{\alpha}{2}}}{n-1}$, but the type I error ($\alpha$) is obtained by searching through numerical methods because of the effect of serial correlation in OTMC.